\documentclass[jcp,aip,preprint,superscriptaddress]{revtex4-1}
\pdfoutput=1
\usepackage{amsfonts}
\usepackage{dcolumn}
\usepackage{graphicx}
\usepackage{amsmath}
\usepackage{amssymb}
\usepackage{siunitx}

\begin{document}

\title{Coating thickness and coverage effects on the forces between
   silica nanoparticles in water}
\date{\today}
\author{K.\ Michael Salerno}
\affiliation{Sandia National Laboratories, Albuquerque, NM 87185, United States}
\author{Ahmed E.\ Ismail}
\affiliation{Aachener Verfahrenstechnik: Molecular Simulations and
Transformations, Faculty of Mechanical Engineering, RWTH Aachen University,
D-52056 Aachen, Germany}
\author{J. Matthew D.\ Lane}
\author{Gary S.\ Grest}
\affiliation{Sandia National Laboratories, Albuquerque, NM 87185, United States}

\sisetup{mode=text,range-phrase = {\text{~and~}}}
\begin{abstract}

The structure and interactions of coated silica nanoparticles have been studied
in water using molecular dynamics simulations. For 5 nm diameter amorphous
silica nanoparticles we studied the effects of varying the chain length and
grafting density of polyethylene oxide (PEO) on the nanoparticle coating's
shape and on nanoparticle-nanoparticle effective forces.  For short ligands of
length $n=6$ and $n=20$ repeat units, the coatings are radially symmetric while
for longer chains ($n=100$) the coatings are highly anisotropic.  This
anisotropy appears to be governed primarily by chain length,
with coverage playing a secondary role. For the largest chain lengths
considered, the strongly anisotropic shape makes fitting to a simple radial
force model impossible.  For shorter ligands, where the coatings are isotropic,
we found that the force between pairs of nanoparticles is purely repulsive and
can be fit to the form $(R/2r_\text{core}-1)^{-b}$  where $R$ is the separation
between the center of the nanoparticles, $r_\text{core}$ is the radius of the
silica core, and $b$ is measured to be between 2.3 and 4.1. 

\end{abstract} \maketitle

\section{Introduction}

Coated nanoparticles have attracted significant industrial and scientific
interest because of their wide range of potential uses, including biomedical
applications such as drug delivery \cite{giljohann.10} and toxin
detection,\cite{uzawa.08} as well as more traditional commercial uses such
as fillers, dispersants, and surfactants in solution.\cite{anselmann.01}
Although the use of nanoparticles continues to increase, our understanding of
the molecular mechanisms through which nanoparticles interact with one another
and with their environments has not kept pace.  It is understood that
nanoparticle coatings can be used to control aggregation and promote solvation
in a processing medium.  Only recently, however, have the effects of variables
such as the length, grafting density, and structure of the polymer chain
coating been directly linked to the dynamics of and interactions between
nanoparticles. \cite{Akcora.2009,Striolo.2007,Rabani.2002}

Experimental study of the stability of nanoparticles in dilute and concentrated
polymer solutions have shown that the length of ligands grafted onto
nanoparticles, compared to the size of the solvent, has a major effect on
stability.\cite{dutta.08} The structure of
brushes, including block copolymers, have been explored with particles produced
by techniques such as reversible addition-fragmentation chain transfer
polymerization.\cite{li.05} Rheometry and dynamic light scattering measurements
have been used to study viscosity and optical effects on aqueous dispersions of
polyethylene oxide (PEO) coated silica.\cite{zhang.04} 
Recent work in this area includes the use of plasmon rulers
to measure interfacial interactions in nanoparticles,\cite{Yoon:2013iza} as
well as low-energy ion scattering \cite{Kauling:2013ix} and infrared
spectroscopy \cite{Andanson:2013gn} to explore the structure of ionic liquids
containing nanoparticles.

The forces and potentials describing the interactions of nanoparticles in
solution have also been studied using computational
approaches.\cite{Likos.2001} For instance, bare colloidal nanoparticles have
been explored in Lennard-Jones \cite{qin.03,qin.06,challa.06} and
\emph{n}-decane \cite{qin.07} solvents, as well as in electrolyte solutions.
\cite{jenkins.07, jenkins.08}  Forces between coated nanoparticles have been
studied in coarse-grained models using Monte-Carlo
methods\cite{Striolo.PRE.2006} as well as in fully atomistic molecular dynamics
(MD) simulations of PbSe nanoparticles capped with aliphatic
chains\cite{Clancy.2012} and silica nanoparticles with PEO
ligands.\cite{lane.09}  MD simulations have also been used to study
nanoparticle monolayers at the surface of monomeric and polymeric Lennard-Jones
liquids,\cite{Cheng:2012hs} and their mechanical properties in a self-assembled
monolayer.\cite{You:2013ec} Larger assemblies of particles have also been
considered. For instance, Lin et al.~\cite{Lin:2011bs} have considered
aggregation in coarse-grained models of functionalized gold nanoparticles.
Other groups have examined the roles of functionalization in silica-polystyrene
systems using both coarse-grained \cite{Ghanbari:2012id} and atomistic
\cite{Ndoro.2011,Ndoro.2012} models. 

There have also been recent theoretical developments in the treatment of
certain nanoparticle systems.  Lin and co-workers have developed a model for
the force required to remove a nanoparticle from a substrate.
\cite{Lin:2013hh} Similarly, mean-field theory has been used to measure how
confinement effects the interactions between assemblies of
nanoparticles.\cite{Pryamitsyn:2013ha} Other work in this area has focused on
the dynamics of large groups of nanoparticles, treating them as coarse-grained
particles using both Brownian dynamics \cite{almusallam.05} and molecular
dynamics.  \cite{karkin.10,udayana.10} Lalatonne et al.\ \cite{lalatonne.04}
have explored the role of dispersion versus dipolar forces in the organization
of magnetic nanoparticles, while, more recently, Pereira and co-workers
\cite{pereira.10} studied ferroelectric nanoparticles embedded in a nematic
liquid crystal.

In our previous work on coated silica nanoparticles \cite{lane.09} we examined
the behavior of functionalized silica nanoparticles approaching either a solid
wall or another nanoparticle.  For the approach of a single particle toward a
wall, the results were shown to be in agreement with the theoretical solution
of Brenner.\cite{brenner.61}  For the approach of two particles, the results
showed that the forces between particles decreases monotonically with
increasing distance. This is in notable contrast to results for bare particles,
which show strong oscillations in the sign and magnitude of the forces between
particles for small separations.\cite{qin.06, challa.06}

Our previous work considered only a single grafting density and chain length in
both equilibrium and non-equilibrium conditions.\cite{lane.09}  In this paper,
we focus solely on equilibrium conditions but extend our analysis to include
the effects of varying the grafting density of ligands, as well as the length
of ligands attached to the surface.  We show that both of these factors affect
the spatial extent of the force interactions between nanoparticles, and combine
to determine the effective radius of the nanoparticle.  In addition, we address
the issue of anisotropy in the structure of coated nanoparticles, showing that
ligand length plays the central role in controlling the level of anisotropy in
the ligand coating.  

Most coarse-grained potential models generally assume that coated spherical
cores will themselves be spherical.  However, for ligand lengths comparable to
the core radius, Lane and Grest have shown previously that asymmetry forms
spontaneously \cite{lane.10} and strongly influences the
nanoparticle-nanoparticle interactions and assembly. \cite{lane.14}  Previous
experimental and simulation results have shown that for low-density or
inhomogeneous graftings anistropic structures can self-assemble.
\cite{Kumar.2013,Akcora.2009}  Our simulations do not probe this regime,
but instead focus on nanoparticle anisotropy in the dense grafting regime.  We
show, in this work, cases where nanoparticle asymmetry produces interactions
which cannot be described via simple radially-dependent potential functions.
We limit our attention in the present work to water, which acts as a good
solvent for PEO.  We have previously reported on the significant effect
solvents can have on the structure of polymer coatings.
\cite{lane.10,Peters.2012} 

We discuss the basic system studied as well as simulation methodology in
Section \ref{secM}.  We present results, including measures of effective
particle radius and particle anisotropy in Section \ref{secRD}, and then
discuss how these measurements relate to and inform our measurements of
interparticle forces.  A summary of findings and some conclusions are presented
in Section \ref{secC}.

\section{Methodology}
\label{secM}

We modeled \SI{5}{nm} diameter amorphous silica nanoparticles which had been
cut from a sample of bulk amorphous silica and annealed to produce a surface
hydroxyl concentration of 4.20 sites per nm$^2$, consistent with
experimental values at 300 K. \cite{zhuravlev.87} The bulk silica was
generated from a melt-quench process similar to the method of Lorenz et al.\
\cite{lorenz.05b}  A passivating coating of methyl-terminated PEO ligands
(Si(OH)$_3$CH$_2$(CH$_2$CH$_2$O)$_n$(CH$_3$)) was then attached via
chemisorption using trisilanol functional groups in place of selected hydroxyl
groups found at the nanoparticle surface.  The chains were added one after
another, each oriented to point radially away from the surface.  If necessary, the
chains were rotated about their long axis until the bulky trisilanol head group
did not overlap the silica core atoms.  Chemisorption sites were selected to be
maximally spaced on the sphere from previously placed chains.

\begin{figure}
\begin{center}
\includegraphics[width=0.8\textwidth, angle=270]{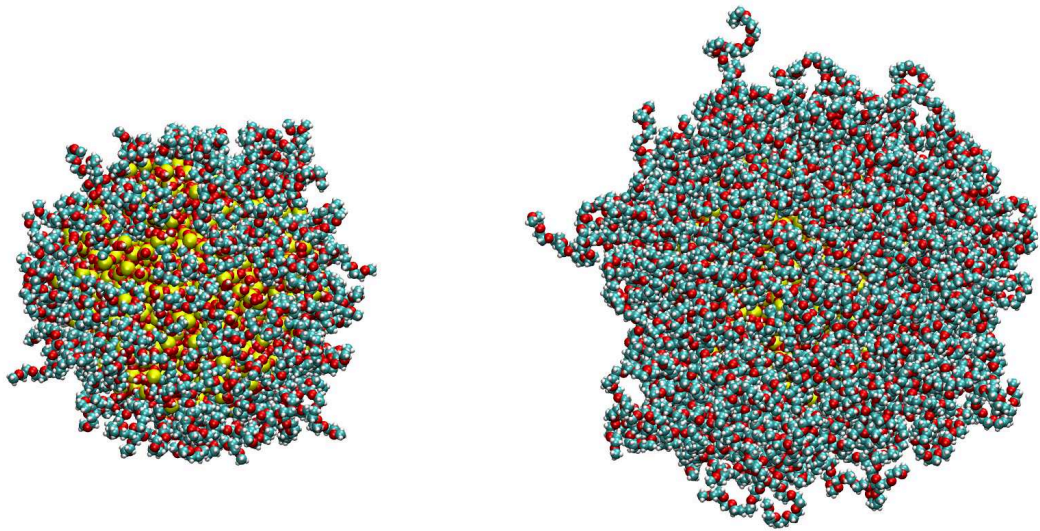}
\end{center}
\caption{Representative view of \SI{5}{nm} diameter silica nanoparticles
functionalized by $n=6 $ (left) and $n=20$ (right) PEO ligands with a grafting
density of \SI{2}{chain/nm^2}.  Water molecules are excluded for clarity.}
\label{f:figure1} 
\end{figure}

The lengths of the ligands studied were $n = 6$, 20, and 100 PEO repeat units.
For $n = 6$, we studied grafting densities $ \sigma = $ 1, 2, and 3
\si{chain/nm^2} or  79, 157, and 234 chains, respectively, per 5 nm-diameter
nanoparticle. For $n = 20$, grafting densities $\sigma = 1$ and
\SI{2}{chain/nm^2} were studied, while for the longest ligands ($n = 100$),
$\sigma =0.2$ and \SI{0.5}{chain/nm^2} were considered, corresponding to 16 or
39 ligands per nanoparticle. These densities are consistent with previous
experimental measurements.\cite{maitra.03}  Examples of silica nanoparticles
with length $n = 6$ and $n = 20$ ligands and grafting density
\SI{2}{chain/nm^2} are shown in Fig. \ref{f:figure1}.

Atom interactions were modeled using all-atom force fields developed by Smith
{\em et al.} for both the PEO \cite{smith.02} and the silica
interactions,\cite{smith.07} and the TIP4P/2005 model for
water.\cite{jorgensen.83, Vega.2005} While Smith {\em et al.} provide explicit
parameters for the PEO-water interactions,\cite{smith.02} the parameters for
the attractive portion of the Buckingham potentials for the interactions
between silica and water were fitted to a Lennard-Jones potential and then
combined with the TIP4P potential using Lorentz-Berthelot mixing rules,
following the methodology of Ref. \citenum{Ismail:2009ky}.

All simulations were carried out using the LAMMPS molecular dynamics
code.\cite{plimpton.95}  The numerical integration was performed using the
velocity-Verlet algorithm with a time step of $\delta t= \SI{1}{fs}.$
Lennard-Jones interactions are truncated at \SI{1.2}{nm}.  Long-range Coulomb
interactions were calculated using the PPPM method with precision
$10^{-4}$.\cite{hockney.88} The nanoparticle core, consisting of the silica
molecules plus the termination of each PEO ligand bonded to the silica, was
treated as a rigid body, while the bond lengths and bond angles of the water
were constrained using the SHAKE algorithm.\cite{ryckaert.77}

To build the nanoparticle-water composite systems, we first
equilibrated a rectangular cell of water at \SI{300}{K} and \SI{1}{atm}
for \SI{1}{ns}. For the $n = 6$ and $n = 20$ nanoparticles, the initial unit
cell size was approximately \SI{13.0}{nm} $\times$ \SI{13.0}{nm} $\times$
\SI{11.5}{nm}; for the $n = 100$ nanoparticles, the unit cell size was
\SI{26.0}{nm} $\times$ \SI{26.0}{nm} $\times$ \SI{23.0}{nm}.  The composite
system was created by inserting the nanoparticle into a spherical hole cut in
the periodic bulk solvent.  The resulting system was then equilibrated 
in the \textit{NPT} ensemble by applying a Nos\'e-Hoover barostat
\cite{frenkel.2001} at \SI{300}{K} and \SI{1}{atm} pressure for at least
\SI{0.5}{ns} with a thermostat time constant of \SI{0.1}{ps} and 
a barostat time constant of \SI{1}{ps}.

Coating structure and anisotropy calculations were made from longer
single-particle simulations in the \textit{NVT} ensemble using a Nos\'e-Hoover
thermostat with time constant \SI{0.1}{ps}.
The center of mass of the rigid core was held fixed during these simulations,
though rotations about the center of mass were permitted.  Atoms in the PEO
coating and water molecules were allowed to move freely.  The configuration of
the particle ligands and the surrounding water was stored every \SI{5}{ps}
during a \SI{4}{ns} sampling period.

Nanoparticle-nanoparticle force calculations were made from two-particle
systems created by replicating the simulation.  After an additional \SI{1}{ns}
equilibration, the particles were brought together by displacing the 
center-of-mass of each core at a rate of \SI{25}{nm/ns} toward the 
center of the box. Positions of all atoms
were stored every \SI{10}{ps}, corresponding to a total net displacement of
\SI{0.5}{nm}. The approach calculation proceeded until the particle cores were
nearly in contact.

The configurations obtained from this ``approach'' simulation were not 
directly used to measure the forces. Instead, we followed the approach
used to produce the static configurations in Ref. \citenum{lane.09}.
Each of the configurations produced from the above run was separately 
simulated using an \emph{NVT} ensemble with the center-of-mass of each
nanoparticle core frozen.
The system was then equilibrated for \SI{1}{ns}, followed by a sampling
run of between \SIrange{3}{6}{ns} per system. 
The force between particles was determined by summing all forces on atoms in
each core whether exerted by the other particle, the surrounding water, or by
the PEO ligands. 
Forces were calculated at each time step and averaged over
\SI{50}{fs} throughout the sampling period. 

\section{Results and Discussion}
\label{secRD}

\subsection{Nanoparticle Structure}

\begin{figure}
\begin{center}
\includegraphics[width=0.9\textwidth]{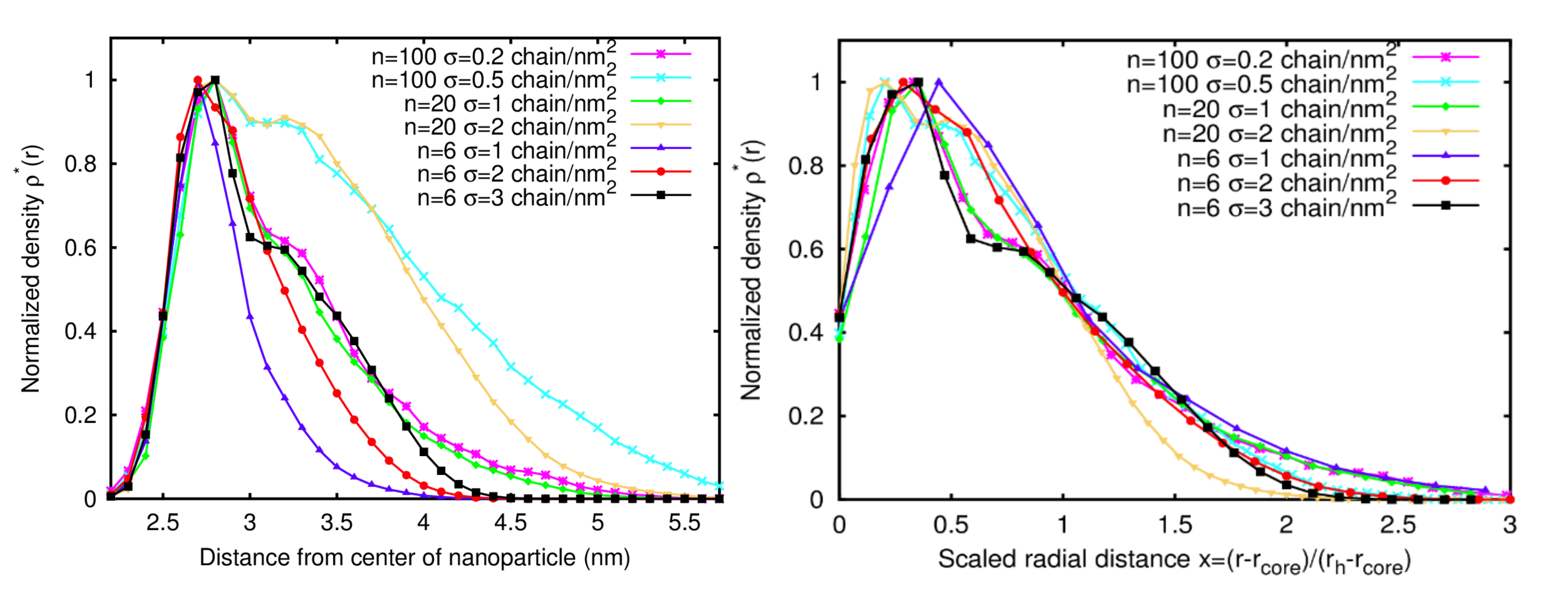}
\end{center}
\caption{(a) Normalized brush density $\rho^*(r) = \rho(r)/\rho_\text{max}$ as a function
of distance from the particle center (in \si{nm}) for ligand length $n = 6$
systems with grafting densities $\sigma = 3$ (black $\blacksquare$), 2 (red
$\bullet$), and 1 \si{chain/nm^2} (blue $\blacktriangle$); for $n = 20$ systems
with densities $\sigma = 2$ (yellow $\blacktriangledown$) and 1 \si{chain/nm^2} (green
$\blacklozenge$); and for $n = 100$ systems for 0.5 (cyan $\times$)
and 0.2 \si{chain/nm^2} (magenta $\ast$).  (b) The same normalized brush density
plotted as a function of scaled distance $x = (r - r_\text{core})/(r_h-r_\text{core}).$ }
\label{singledens} 
\end{figure}

Examining first the PEO ligand structure, we construct a radial mass density
profile $\rho(r)$ for the PEO ligands as a function of distance from the center
of mass of the silica particle.  The maximum density, $\rho_\text{max}$ is used
to normalize the density profile.  Figure \ref{singledens} (a) shows the
resulting normalized density profiles $\rho^*(r) = \rho(r) / \rho_\text{max}$
for different brush lengths and grafting densities.  From the data in this
figure we estimate the half-maximum radius $r_h$ as the largest distance at
which the normalized density $\rho^*(r_h) = 0.5$.  As in previous studies of
nanoparticle ligands in good solvents,\cite{Peters.2012} we find that the
ligand density is a good measure of the ligand extension into the solvent.
There is a notable shoulder in all curves except for the $n = 6$ chains at
low coverage $\sigma = $1 - \SI{2}{chain/nm^2}.  This shoulder corresponds to
the first carbon - oxygen sequence along the PEO backbone, with its increased
mass density.  The absence of the feature from the short, low-density
configurations indicates that in these systems the PEO chains are able to
orient more freely, even very near to the surface.

Examining the half-maximum radii shown in Table \ref{t:radii} we see that $r_h$
increases monotonically with increasing coverage density for a given ligand
length. This can be explained by the radial orientation of the chains caused
by packing effects.  Comparison of values between different ligand
lengths, however, suggests that it is the total number of repeat units in the
coating that determines the half-maximum radius: note that the half-maximum
radius remains unchanged when changing between configurations with
approximately the same total number of repeat units, such as $n = 20$,
$\sigma = \SI{1}{chain/nm^2}$ and $n = 100$, $\sigma = \SI{0.2}{chain/nm^2}$ or
$n = 20$, $\sigma = \SI{2}{chain/nm^2}$ and $n = 100$, $\sigma =
\SI{0.5}{chain/nm^2}$.

\begin{table}[bt]
\caption{ Characteristic radii $r_h$ and $r_\text{max}$, relative 
standard deviations of the grafting, chain, and methyl densities as
a function of ligand length and grafting density, and ratio of largest 
to smallest moment of inertia eigenvalues}

\begin{tabular}{cccccccc} \hline\hline

Ligand length & $\sigma $ & $r_h$ & $r_\text{max}$ & 
\multicolumn{3}{c}{Relative standard deviation (\%)} & Eigenvalue Ratio\\
\cline{5-7} 
$n$ & (\si{chain/nm^2}) & (\si{nm}) & (\si{nm}) & 
Grafting density & Coating density & Methyl density & $\lambda_{max}/\lambda_{min}$ \\
\hline
6 & 1 & 2.95 & 4.35 & 10.5 & 10.3 & 17.0 & 1.15 \\
6 & 2 & 3.20 & 4.35 & 10.2 & 8.6 & 10.1 & 1.12 \\
6 & 3 & 3.35 & 4.65 & 8.6 & 6.3 & 8.1 & 1.10 \\
20 & 1 & 3.35 & 5.75 & 10.4 & 13.4 & 18.5 & 1.21 \\
20 & 2 & 3.95 & 5.95 & 10.3 & 8.7  & 15.1 & 1.19 \\
100 & 0.2 & 3.35 & 5.30 & 30.7 & 28.3 & 59.8 & 1.70 \\
100 & 0.5 & 3.95 & 6.50 & 15.4 & 25.5 & 30.4 & 1.75 \\ \hline\hline

\end{tabular}
\label{t:radii}
\end{table}

Another measure of ligand conformation is the radius of maximum extent
$r_\text{max}$, which measures the distance the ligands extend beyond the
nanoparticle surface.  Values of $r_\text{max}$ are reported in Table
\ref{t:radii}.  For the $n = 6$ nanoparticles, the maximum extent of the ligands is
between \SIrange{1.8}{2.2}{nm}. Given that the backbone end-to-end length is
approximately \SI{2.2}{nm}, this implies that the ligands are largely uncoiled,
with a greater fraction uncoiled at higher grafting densities. For the $n = 20$
system the ligands extend between 2.3 and 2.9 \si{nm}.  This is somewhat larger
than the expected end-to-end distance of a random coil, but much shorter than
the distance along the uncoiled backbone. This is consistent with the ligands
near the particle core being partially ordered, with a transition to a random
coil-like orientation further from the core, closer to the coating surface.
Finally, for the $n = 100$ ligands, the maximum extent is comparable in
magnitude to the end-to-end distance in solution.\cite{Smith.2000}  This
suggests that for longer ligands with low grafting densities the ligands may be
able to attain a coil-like orientation in space.  At higher densities, however,
there may be interactions between the chains, leading to slightly more ordered
chains and thus a larger maximum extent.

From the half-maximum radius we may define a dimensionless radial coordinate
\begin{equation}
 x = \frac{(r - r_\text{core})}{(r_h - r_\text{core})},
\end{equation}
where $r_\text{core} = \SI{2.50}{nm}$ is the radius of the silica core.  This
coordinate can be used to compare the PEO density profiles across different
ligand length and coverage configurations.  Figure \ref{singledens} (b) shows
the normalized PEO density $\rho^*(r)$ as a function of $x$.  
The coordinate $x$ captures much of the
variation in radial ligand density for different chain configurations, although
there is some variation in the scaled curves.  The location of the density
maximum moves closer to the particle surface as $r_h$ increases. This is
expected from the nearly uniform density maximum in Fig.  \ref{singledens} (a)
and results in part from the choice of $r_\text{core}$. For $x > 2$ the tail of
the density profile is systematically depressed with increasing grafting
density.  This variation highlights the limitations of describing the complex
interplay of the competing grafting density and chain length effects with the
simple $r_h$ length scale.

\begin{figure}
\begin{center}
\includegraphics[width=4.30in]{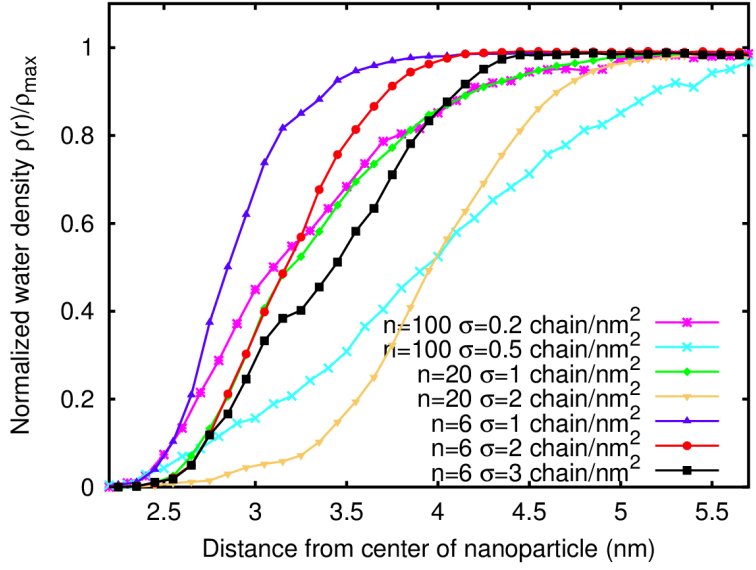}
\end{center}
\caption{Normalized water density as a function of distance from the
particle center (in nm) for chain length $n = 6$ systems for grafting
densities of 3 (black $\blacksquare$), 2 (red $\bullet$), and 1 \si{chain/nm^2}
(blue $\blacktriangle$); for $n = 20$ systems for densities of 2
(yellow $\blacktriangledown$) and 1 (green $\blacklozenge$) \si{chain/nm^2}; and for $n =
100$ systems for 0.5 (cyan $\times$) and 0.2 (magenta $\ast$) \si{chain/nm^2}.
}
\label{waterdens} \end{figure}

We also examine the density of water as a function of distance from the
nanoparticle center.  In our previous simulations of PEO and water on a flat
substrate, \cite{ismail.07} we have shown that water penetrates readily even
into relatively dense polymer brushes. Other simulations of gold nanoclusters
with PEO groups along the ligands found that water readily penetrates into such
ligands.\cite{Yang.2011}  Thus, a comparison of the water density as a function
of radius versus the bulk density can provide information regarding the
structure of the PEO coating around the silica core.  Figure \ref{waterdens}
shows the density of water molecules as a function of distance to the
nanoparticle center of mass.

From Fig.  \ref{waterdens} we see that, as for the half-maximum radius, the
total number of repeat units grafted onto the particle appears to be the most
important factor in determining the water density profile: for instance, there is
relatively little difference between the systems with $n = 6$ at
\SI{3}{chain/nm^2}, $n = 20$ at \SI{1}{chain/nm^2}, and $n = 100$ at
\SI{0.2}{chain/nm^2}, containing approximately the same total number of repeat
units.  The primary difference between the curves is that the radius beyond
which the density of water equals its bulk density increases with $n$.  

\subsection{Particle anisotropy}

In some cases, a nanoparticle's coating ligands are observed to orient and
close pack together.  This ordering of the chains can be observed in spatial
variations of the local coating densities.  We refer to these spatial
variations and the oriented ligands that cause them as characteristics of
anisotropic coatings.  We describe, here, a method for quantifying the degree
of anisotropy in coatings and show later that, in its extreme, this asymmetry
can destroy the radial symmetry of the interaction forces.

We characterize the surface inhomogeneity of each particle's coating by the
standard deviation in the local chain density over the surface. The local
density of PEO chain atoms was calculated at 2000 overlapping patches equally
spaced over the sphere's surface.\cite{Clark:1978wy} An atom was determined
to be in an angular patch about a point if it was inside a cone with vertex at
the nanoparticle center and axis passing through said point on the surface.
Cones with \ang{90} opening angles were used, giving each patch an area of
\SI{11.5}{nm^2} at the surface, or approximately \SI{15}{\%} of the total
surface area of the nanoparticle. This patch size was selected because it is a
reasonable contact patch for two approaching particles.  From the collection of
patch densities a mean and standard deviation can be calculated.  The mean
simply returns the overall density of the coating.  The standard deviation is a
measure of how this density varies locally on the surface, a measure of
asymmetry.  In order to compare across particles with different grafting
densities, we report the relative standard deviation i.e.\ the standard deviation
as a percentage of the mean.  This calculation was performed for the grafting
points as well as the total coating and the terminal methyl groups.

The relative standard deviation (RSD) of the local coating, grafting point, and
methyl group densities are shown in Table \ref{t:radii} for the different
particles studied.  Several trends are evident.  First, the relative standard
deviation in the grafting density decreases slightly with increasing grafting
density, as more densely packed coatings are less likely to have gaps in the
coatings which increase the RSDs of the patch densities.  We see that for short
chains the coating asymmetry is somewhat lower than the grafting asymmetry.
However, for lower grafting density and longer chains both the coating
asymmetry and the methyl asymmetry can be significantly higher than the
grafting asymmetry.  This indicates that long chains orient and close pack more
strongly than shorter chains and this leads to more anisotropic coatings.  This
effect is likely to be dramatic for coating ligands in a poor solvent, leading
to even greater anisotropy.

Particle anisotropy can also be quantified using the eigenvalues of the moment
of inertia tensor.  The ratio of the largest to smallest eigenvalues compares
the smallest and largest axes of the particle, giving a sense asymmetry or
anisotropy.  The ratio will equal one for spherically symmetric objects, while
it will be larger than one for objects which are not spherically symmetric.
Eigenvalue ratios for the different particles studied are listed in Table
\ref{t:radii}.  The trend of increasing anisotropy with increasing chain length
is clearly reflected.  Particles with higher grafting density also tend to be
slightly more isotropic, as expected, though this trend breaks down for the
most anisotropic particles.

\begin{figure}[tb]
\begin{center}
\includegraphics[width=3.8in]{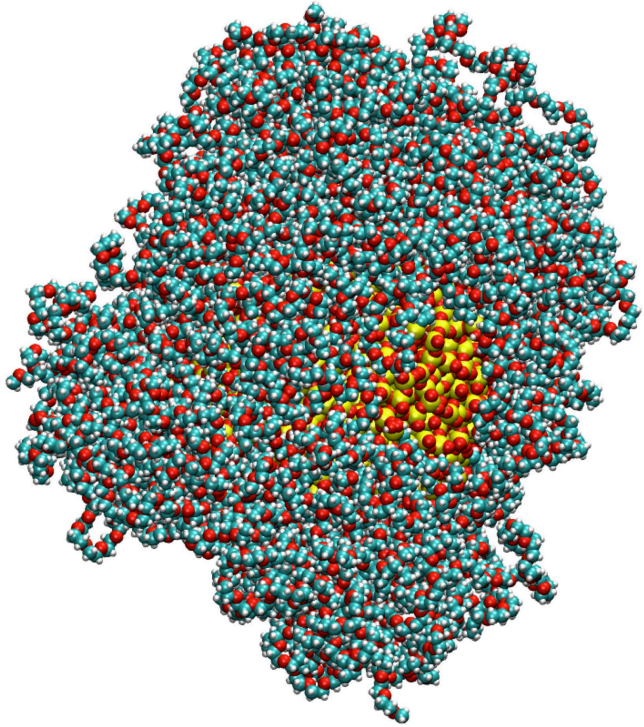}
\end{center}
\caption{Sample image of a \SI{5}{nm} silica nanoparticle, coated with
PEO of 100 repeat units at a grafting density of \SI{0.5}{chain/nm^2}.}
\label{peo5100}
\end{figure}

An example of a highly anisotropic particle can be seen for the \SI{5}{nm}
silica nanoparticle coated by $n = 100$ chains at grafting density
\SI{0.5}{chain/nm^2} shown in Fig. \ref{peo5100}. The anisotropy of the grafted
ligands for this particle contrasts with the spherically symmetric particles
shown in Fig. \ref{f:figure1}.  For asymmetric particles, the function
representing the force between particles is no longer a function only of the
interparticle distance, since the effective radii of the particles may vary
significantly with direction.  For this reason, the $n=100$ nanoparticles were
excluded from the analysis of forces below.  Experimental measurements of PEO
grafted onto silica nanoparticles find that for long chains surface densities
consistent with our most anisotropic particles are possible i.e.\ $\sigma =
\SI{0.2}{chain/nm^2}$ for $n \approx 100,$ \cite{maitra.03} indicating that
such anisotropic coatings may be found in experimentally realizable systems. 

\subsection{Forces between nanoparticles}
\label{sec:force}

Using the two-particle systems described in Section \ref{secM} we measure the
forces between particles at a number of radial separations for the $ n = 6 $
and $ n = 20 $ systems.  As described above, we limit our analysis to these
systems because we assume a spherically symmetric force, which is not a valid
assumption for the longer chains.  As expected, we find that the interparticle
forces decrease as a function of radial separation $R$.  At small separations
interparticle forces rapidly exceed those found in equilibrium systems or
systems under shear.  The location of the divergence is set by the nanoparticle
radius, while the scale of the interaction depends on both the chain length $n$
and the grafting density $\sigma$.  We can test whether different nanoparticle
configurations collapse on a single force curve using the scaling based on the
half-maximum radius.

For all chain lengths we find that the forces are purely repulsive and depend
strongly upon the grafting density.  Similarly, we find that for a fixed
grafting density, the interparticle force also depends on chain length.  For
particles with a low grafting density of short chains ($n=6$, $\sigma =
\SI{1}{chain/nm^2}$), the forces are below \SI{1}{nN} even when there is only a
\SI{0.5}{nm} gap between the silica surfaces of the particles.  For chains with
$n = 20$ at higher grafting density, the interparticle forces can remain
non-zero out to particle separations greater than \SI{9}{nm}.  For both short
and long chain particles the magnitude of the equilibrium force appears to fall
off to zero at a distance between $2r_h$ and $2r_\text{max}$.

\begin{figure}[tb]
\begin{center}
\includegraphics[width=4.3in]{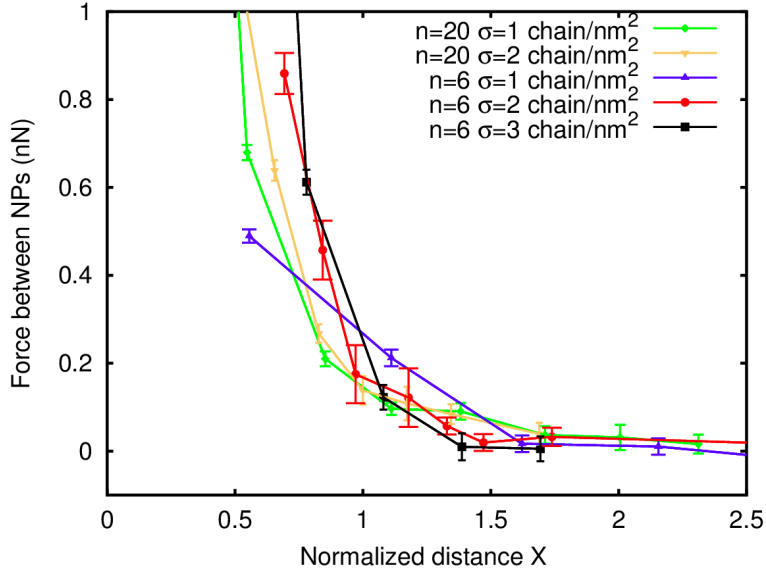}
\end{center}
\caption{ Equilibrium forces between functionalized nanoparticles as a function
of shifted and scaled radial separation $X = (R - 2r_\text{core})/2(r_h -
r_\text{core})$ for chain length $n = 6$ systems for grafting densities of 3
(black $\blacksquare$), 2 (red $\bullet$), and 1 \si{chain/nm^2} (blue
$\blacktriangle$); and for $n = 20$ systems for densities of 2 (yellow
$\blacktriangledown$) and 1 (green $\blacklozenge$) \si{chain/nm^2}.  }
\label{f:normforces} 
\end{figure}

Additional insight can be gained from plotting the force curves
against
\begin{equation}
  X = \frac{(R - 2r_\text{core})}{2(r_h - r_\text{core})},
\end{equation}
the distance between the surfaces of the two particles normalized by the width
of both brushes, as shown in Fig. \ref{f:normforces}.  When this scaling is
taken into account, it appears that all of the systems follow the same trend:
large forces for $X < 1$, followed by forces approaching zero between $ 1 < X <
2$.  There is some variation between the scaled curves, indicating that the
interparticle force is not entirely governed by half-maximum radius $r_h.$  In
particular, the system with the fewest total repeat units, $ n = 6$, $ \sigma =
\SI{1}{chain/nm^2} $ appears to have a much softer force curve.  For the denser
$ n=6$ and $ n = 20 $ systems the scaled force curves agree and rapidly
increase as $X$ is reduced from roughly \num{1.5} to \num{0.5}.

\begin{figure}[tb]
\begin{center}
\includegraphics[width=4.3in]{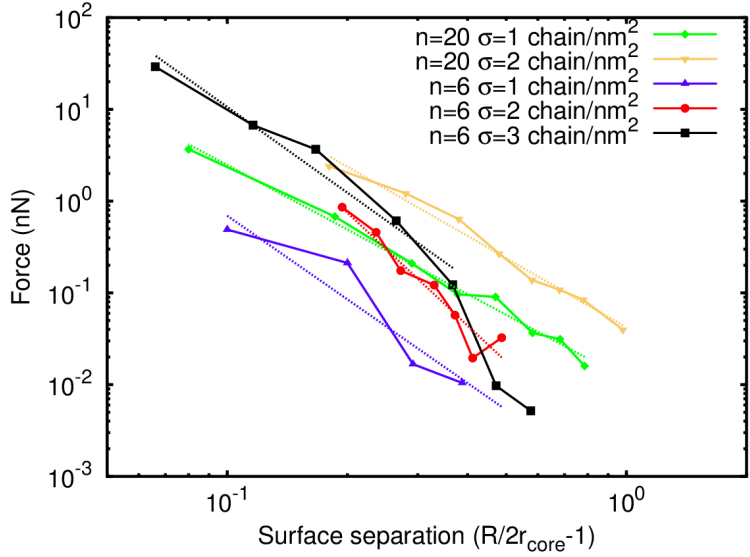}
\end{center}
\caption{Force curves as a function of shifted particle separation
$(R/2r_\text{core}-1)$ for the studied systems plotted on log-log axes.  Fits
are shown as dashed lines, with parameters listed in Table \ref{t:2v}.  }

\label{f:llforces} 
\end{figure}

Shifted force curves can be plotted on log-log axes in order to compare the
steepness of the slope increase.  Because a divergence is expected at radial
separation $R=\SI{5.0}{nm}$, the curves have been shifted by the nanoparticle
core diameter.  These curves are shown in Fig. \ref{f:llforces} and can be fit
to the form of a prefactor times a power of the surface separation $F(R) =
A(R/2r_\text{core}-1)^{-b}$.  For the particles studied, we find
$\SI{6.5e-4}{nN} < A < \SI{4.3e-2}{nN}$ and $2.3 < b < 4.1$.  These fits are
shown as dashed lines in Fig \ref{f:llforces}, with parameters recorded in
Table \ref{t:2v}.  Although the range of data is small, the trend indicates
that the slope is steeper at smaller ($n = 6$) chain lengths, fitting the
intuitive picture that short chains lead to stiffer interactions.  The trend is
also consistent with the geometric picture of longer chains containing both
coiled and uncoiled conformations, leading to a softer interaction.

\subsection{Second Virial Coefficient}

The second virial coefficient is related to the osmotic pressure of a
collection of macromolecules in solution, and is directly accessible via
multiple experimental measurements.  The numerical value of the coefficient can
also be calculated from the force or potential between the nanoparticles
measured during simulation.\cite{Likos.2001} From this calculation we give an
estimate of the value of the second virial coefficient for our nanoparticles
and compare these estimates with experimental values.  

The standard calculation of the osmotic second virial coefficient $A_2$ is based on
the definition 
\begin{equation} 
	A_2 = \frac{2\pi N_A}{M^2} \int_0^\infty (1-e^{-U(R)/kT})
	R^2\,dR \label{eq:2v} 
\end{equation} 
where $U(R)$ is the potential between two interacting particles, $M$ is the
molar mass of a nanoparticle, and $N_A$ is Avogadro's number.  The definition above
depends only on the radial distance $R.$  As before, this approximation is valid
only for the $n = 6$ and $n = 20$ isotropic particle cases.  We calculate the
second virial coefficient based on the forces measured in Section
\ref{sec:force}, and determine the interparticle potential as the integral of
the force with respect to radial distance $R$.  The force can be integrated
analytically to produce a potential $U(R) =
C_1/(R/2r_\text{core}-1)^{b-1}+C_0,$ where $C_0$ is chosen so that the
potential goes continuously to zero beyond an upper radial cutoff.  The
potential is considered to be infinite for $R < 2r_\text{core}$ and zero for
$R$ larger than the upper cutoff.

For the nanoparticle systems studied we need to estimate an upper radial cutoff,
beyond which both the force and potential are zero.  One estimate $R_0$ for
the radial cutoff is the point at which the magnitude of the fluctuations in the
force are larger than the magnitude of the force itself, so that the force is
effectively zero.  This occurs for forces less than about \SI{0.05}{nN} and the
large fluctuations below this level can be seen on both the linear and log
scales in Figs.  \ref{f:normforces} and \ref{f:llforces}.  Another, more
conservative estimate of the radial cutoff is $2r_\text{max}$.  Because the
interparticle forces are caused by steric interactions, the largest particle
separation that can contribute to the integral is $R=2r_\text{max}$.  Based on
these cutoff radii and the potential $U(R),$ numerical values of the integral in
Eq. \ref{eq:2v} can be calculated.  Both fit values and estimates of the second
virial coefficient are recorded in Table \ref{t:2v}.  We note that using
$2r_\text{max}$ as an upper cutoff assumes that the functional form of the
force and potential do not change beyond where they can be reliably measured
with molecular dynamics.  Checking the validity of this assumption is a very
costly computational task, but the variation of the exact cutoff value does not
appear to dramatically change the calculated value of the second virial
coefficient.

The osmotic second virial coefficient has been experimentally measured in a
number of sterically stabilized nanoparticle systems. Jansen \textit{et al.}
studied silica particles of radius \SI{31}{nm} with linear carbon chain ligands
in toluene, finding second virial coefficients at room temperature similar in
magnitude to those reported here.\cite{Jansen.1986}  More recently, the second
virial coefficient of CdSe nanocrystals with grafted trioctylphosphine oxide
was measured using membrane osmometry.\cite{Striolo.2002}  For comparable
particle sizes (\SI{4.6}{nm} diameter) measurements of the virial coefficient
varied between \num{1.5e-5} and \SI{7e-5}{cm^3 mol/g^2}, in line with our
measurements.  Measurements on slightly different nanoparticle solvent systems
also give measurements of similar magnitude, about \SI{2e-5}{cm^3
mol/g^2}.\cite{Mackay.2006}

\begin{table*}[bt]
\caption{Fit parameters and values of the osmotic second virial coefficient
for different nanoparticle-ligand configurations.}
\begin{tabular}{cccccc} \hline\hline
$n$ & $\sigma$ & $A$ & $b$ & $A_2$ from $R_0$  & $A_2$ from $2r_{\max}$  \\
 & (\si{chain/nm^2}) & (\si{$\times 10^{-4}$}{nN}) & & 
 (\SI[retain-unity-mantissa=false]{1e-5}{cm^3 mol/g^2}) 
 & (\SI[retain-unity-mantissa=false]{1e-5}{cm^3 mol/g^2}) \\
\hline
6 & 1 & 6.5 & 3.0 & 2.1 & 3.8 \\
6 & 2 & 11 & 4.1 & 2.0 & 3.2 \\
6 & 3 & 85 & 3.1 & 2.0 & 3.3 \\
20 & 1 & 110 & 2.3 & 2.0 & 5.6 \\
20 & 2 & 430 & 2.5 & 2.0 & 4.4 \\ \hline\hline
\end{tabular}
\label{t:2v}
\end{table*}

\section{Conclusions}

\label{secC}

We have explored the structure of PEO-coated silica nanoparticles as a function
of chain length and grafting density, as well as the interactions between pairs
of particles. The total mass grafted onto the surface appears to be the most
important factor in determining the structure of the particle, as both the
density profile and half-maximum radius appear to vary directly with the total
number of repeat units. The maximum extent of the chains from the particle
surface also increases weakly with grafting density. This suggests that 
long chains undergo a transition from a primarily linear structure near the
grafting point to a more coil-like structure far away from the grafting point.

The force between two nanoparticles was determined as a function of separation
for all nanoparticle configurations. As chain size increases, the tendency of
the chains to achieve a coil-like conformation increases, influencing the extent
of the chains and the half-maximum radius $r_h.$  As chain length increases the
anisotropy of the particle also increases to such an extent that the
assumption of a radially symmetric particle breaks down, and the relative
orientation of the nanoparticles needs to be considered in describing the
interactions between them.  When particles are separated by less than twice the
half-maximum radius $r_h$ of a particle, the forces rapidly increase; for
separations greater than about $2.5r_h$, it is difficult to distinguish
interactions beyond statistical uncertainty.

The second virial coefficient was calculated from analytical fits of the
measured interparticle forces and assumption of a radial cutoff.  Coefficient
values fell in the range of 1-10 $\times 10^{-5}$ cm$^3$ mol/g$^2$ and are
comparable to experimental measurements.  For the purely repulsive forces found
here, this coefficient represents and effective nanoparticle volume due to
the steric interactions of the PEO chains.

\section*{Acknowledgments}

AEI was supported by the Cluster of Excellence ``Tailor-Made Fuels from
Biomass,'' which is funded by the Excellence Initiative by the German federal
and state governments to promote science and research at German universities.
This work was performed, in part, at the Center for Integrated Nanotechnology,
a U.S. Department of Energy and Office of Basic Energy Sciences user facility.
Sandia National Laboratories is a multi-program laboratory managed and operated
by Sandia Corporation, a wholly owned subsidiary of Lockheed Martin
Corporation, for the U.S. Department of Energy's National Nuclear Security
Administration under contract DE-AC04-94AL85000. 

\bibliographystyle{apsrev4-1}
%\bibliography{nano2}
%
\end{document}